\documentclass[pre,twocolumn]{revtex4}

\usepackage{graphicx}% Include figure files
\usepackage{epsfig}
\usepackage{bm}% bold math
\usepackage{amsmath}
\usepackage{amssymb}
\usepackage{amsfonts}
\usepackage[usenames]{color}
\usepackage[normalem]{ulem}
\begin{document}

\title{Nonlinear Schr\"{o}dinger Equation with Random Gaussian Input: \\
Distribution of Inverse Scattering Data and Eigenvalues}

\author{Pavlos Kazakopoulos}
% \email{pkazakop@phys.uoa.gr}
\author{Aris L. Moustakas}
% \email{arislm@phys.uoa.gr}
\affiliation{Department of Physics, University of Athens, Athens
15784, Greece.}

%\date{\today}

\begin{abstract}
We calculate the Lyapunov exponent for the non-Hermitian
Zakharov-Shabat eigenvalue problem corresponding to the attractive
non-linear Schr\"{o}dinger equation with a Gaussian random pulse
as initial value function. Using an extension of the Thouless
formula to non-Hermitian random operators, we calculate the
corresponding average density of states. We also calculate the
distribution of a set of scattering data of the Zakharov-Shabat
operator that determine the asymptotics of the eigenfunctions. We
analyze two cases, one with circularly symmetric complex Gaussian
pulses and the other with real Gaussian pulses. We discuss the
implications in the context of information transmission through
non-linear optical fibers.
\end{abstract}

\maketitle

\section{Introduction}

One  defining development in telecommunications technology during
the last two decades has been the widespread use of optical fibers
for transmitting enormous quantities of data across large -- even
transoceanic -- distances. For such increasingly large distances,
the non-linearities in the fiber
cannot be neglected, as they tend to distort
transmitted pulses. Consequently, the detection of
traditionally modulated signals becomes problematic.
 For fibers with negative group velocity
dispersion (GVD) it is possible to compensate these effects by
creating stable solitonic pulses
\cite{Hasegawa_Tappert_1973_anomalous,
Mollenauer_Stolen_Gordon_1980}.  As a first approximation, these
 solitary waves are solutions of
the non-linear Schr\"{o}dinger equation (NLSE), the effective
equation describing propagation of light
 in the frame comoving with the mean group
velocity \cite{Agrawal_1995}. In normalized units the NLSE is
expressed as
\begin{equation}
\label{nlse}
i\frac{\partial u}{\partial x} + \frac{\partial^2 u}{\partial t^2
}+2|u|^2u = 0
\end{equation}
where $u(t,x)$ is the (complex) envelope of the electric field,
carrying the transmitted information signal along the
fiber \footnote{Note that in the analysis of light
propagation through optical fibers the roles of space and time have
been exchanged, as compared to the traditional nonlinear
Schr\"{o}dinger equation: Propagation down the fiber has taken the
place of time, while the traditional space variable has been
replaced by the retarded time measured in the frame moving along
the fiber with the group velocity.}. Traditional analyses of
this equation focus on single and dilute solitonic
propagation \cite{Essiambre1997_TimingJitterSolitons1}. However,
to address the ultimate information capacity limits through
the fiber using solitonic pulses, one needs to explicitly consider
dense soliton systems, where the soliton interactions can no
longer be treated as small.

The problem of determining the spatial evolution of an incoming
pulse $u(t)\equiv u(t,0)$ is solved via the inverse scattering
transform (IST), where $u(t)$ enters as the ``potential" in a
linear eigenvalue problem. For the NLSE this is the
Zakharov-Shabat (ZS) eigenvalue problem
\cite{Zakharov_Shabat_1972}, comprising of a $2\times 2$ system of
coupled first order differential equations, %
\begin{eqnarray} %
 \left( \begin{array}{lr} i\partial_t &
u^*(t)
\\ -u(t) & -i\partial_t \end{array}
\right) \bm{\Psi}_z(t) \equiv  \bm{U}(t) \bm{\Psi}_z(t)= z\bm{\Psi}_z(t), %
\label{zs}
\end{eqnarray} %
where $\bm{\Psi}_z(t)=\left[ \psi_1(t) \,\, \psi_2(t)  \right]^T$, and
appropriate asymptotic conditions on the eigenstates, given in the next section.

In this paper we analyze the distribution of the scattering data,
\emph{i.e.} the average density of states (DOS) of $\bm{U}$ and
the average distribution of a set of complex numbers $\{b_z\}$
that determine the asymptotics of the eigenstates,
when $u(t)$ is drawn from a zero-mean, $\delta$-correlated
Gaussian distribution, describing the distribution of transmitted
codewords. Gaussian input signals are often used in information
theory, and in linear transmission problems they often reach the
Shannon capacity \cite{Cover_Thomas_book}. In addition, when the
characteristic signal amplitude $u_0$ is much smaller than its
bandwidth $\tau^{-1}$ (but with $D\equiv u_0^2\tau$ arbitrary), it
is reasonable to approximate \cite{Gredeskul1990_Dark_Solitons} the
input distribution with a $\delta$-correlated Gaussian for
eigenvalues $z$ small in the scale of $\tau^{-1}$.

The non-hermiticity of $\bm{U}$ causes the eigenvalues to spread
over the complex plane. This generally makes the exact calculation
of the DOS more difficult. Several powerful methods have been
developed for calculating the statistical properties of
non-Hermitian operators, which appear in the modelling of diverse
physical processes (see  \emph{e.g.} \cite{Di_Francesco_1994,
Stephanov_1996, Forrester_Jancovici_1996,
Feinberg_Zee_Hermitization, Miller_Wang_1996, Chalker_Wang_1997,
Janik_April_1997, Janik_June_1997, Biane_Lehner_2001,
Fyodorov_Sommers_2003, Wiegmann_Zabrodin_2003,
Gudowska-Novak_et_al, Fyodorov_Khor_Sommers_1997}). In most
cases the random matrices are treated in a mean-field sense and
are thus considered full random matrices. However, to our
knowledge there are only few non-Hermitian
operators with diagonal randomness
for which the exact density of states has been
calculated in closed form \cite{Hatano_1998_review,
EBrezinAZeeNuclPhB509_1998_p599,
Goldsheid_Khoruzhenko_1998_Anderson_models}. In our case, we first
calculate the Lyapunov exponent in closed form taking advantage of
its self-averaging properties. Combining this with a
generalization of the Thouless formula \cite{Thouless_1972} for
non-Hermitian operators
\cite{IYGoldscheidBAKhoruzhenko_ThoulessNHermitian}, that relates
the Lyapunov exponent with the DOS, we arrive at an explicit
expression for the latter. Since the Lyapunov exponent is simply
related to the localization length, it also provides information
for the eigenfuctions of $\bm{U}$.

In addition to the DOS we calculate the limiting distribution
of the scattering data coefficients $\{b_z\}$, which depends
strongly  on the input distribution of $u(t)$: For circularly
complex $u(t)$ the distribution of $\ln b_z$ approaches a Gaussian
distribution albeit with singular variance growing as $T\ln{T}$,
while for real $u(t)$ the distribution is highly singular,
approaching a Cauchy distribution.

It should be noted that the Hermitian ``counterpart" of this
operator,
\begin{equation}
\bm{U}_H=\left(
\begin{array}{lr} i\partial_t & u^*(t) \\ u(t) &
-i\partial_t \end{array} \right)
\end{equation}
arises in the IST for positive GVD, and also as a special case of
the fluctuating gap model of disordered Peierls chains (see
\cite{Bartosch_thesis} and references therein). Its DOS and
localization length have a long history of analysis
\cite{Ovchinnikov_Erikhman_1977, Hayn_John_1987,
Gredeskul1990_Dark_Solitons, Bartosch_Kopietz_1999_Numerical_DOS}.

The spectrum of $\bm{U}$, together with the asymptotic behavior of
the corresponding eigenstates $\bm{\Psi}_z$, which as we
shall see is determined by $b_z$, have the same information
content as the input signal $u(t)$. This is because inverse
scattering transform mapping between the scattering data of all
eigenstates and $u(t)$ is one-to-one \cite{Konotop_book}
\footnote{In the case of a single localized eigenstate with $z=\xi
+ i\eta$ and $b=|b|e^{i\phi}$, the corresponding soliton has
amplitude $\xi$, velocity $2\eta$, initial ``position''
$t_0=\frac{\ln |b|}{2\eta}$ and initial phase $\phi_0=\phi-\pi$
\cite{Konotop_book}.}.  However, while the
spatial evolution of $u(t,x>0)$ and the eigenstates
${\bm{\Psi}(t,x>0)}$ is quite complicated,
the eigenvalues $z$ of $\bm{U}$ remain constant as the signal
propagates down the fiber, and the corresponding
scattering data vary in a trivial manner \cite{Konotop_book}. In
fact, they can both be seen as playing the role of ``action''
variables changing adiabatically in the presence of non-integrable
perturbations. Therefore, the problem of light propagation in the
fiber becomes easier to analyze in terms of the scattering data of
the Zakharov-Shabat eigenproblem, especially in the presence of
perturbations to (\ref{nlse}), such as noise due to amplification
or phase conjugation, which will ultimately determine the optical
fiber capacity \cite{Mitra_Stark_2001, Green_et_al_2002_XPM,
Turitsyn_et_al_2003_zero_average_dispersion,
Kahn_Ho_2004_IT_review}. As a result, the description of the
scattering data as a function of the input signal $u(t)$ may
provide a framework for understanding the ultimate limits of
information transfer through optical fibers.

\section{Lyapunov Exponent and DOS}

We will now describe the basic steps to calculate the
Lyapunov exponent of ${\bm U}$ in (\ref{zs}) which will then lead
to the average DOS. To proceed, we start by introducing the ZS
eigenvalue problem. Traditionally, this is defined as a scattering
problem of the operator ${\bm U}$ in (\ref{zs}), in the presence
of the potential $u(t)\equiv u(t,x=0)$, which decays sufficiently
fast for $t\rightarrow\pm\infty$. In this context the scattering
states are set up with the following asymptotic conditions outside
the range of the potential:
\begin{eqnarray} %
 \begin{array}{l} \bm{\Psi}_z(t) \rightarrow \left ( \begin{array}{c} 0 \\ 1 \end{array} \right )
  e^{izt},\;\; \bar{\bm{\Psi}}_z(t) \rightarrow \left ( \begin{array}{c} 1 \\ 0 \end{array} \right )
  e^{-izt} \;\;\;\; \mbox{as} \; \; t \rightarrow \infty \\ \\
  \bm{\Phi}_z(t) \rightarrow \left ( \begin{array}{c} 1 \\ 0 \end{array} \right )
  e^{-izt}, \; \bar{\bm{\Phi}}_z(t) \rightarrow \left ( \begin{array}{c} 0 \\ 1 \end{array} \right )
  e^{izt} \;\;\;\; \mbox{as} \; \;  t \rightarrow -\infty. \end{array}%
\label{zs_asymptotic}
\end{eqnarray} %

For concreteness, we express the eigenvalue $z$ as $z =
\xi+i\eta$. The two sets of
solutions in (\ref{zs_asymptotic}) are linearly related
through the $S$-matrix:
\begin{eqnarray}
\left[\begin{array}{c} %
\bm{\Phi}_z(t) \\
\bar{\bm{\Phi}}_z(t) \end{array} \right] =
\left(\begin{array}{cc} %
b(z) & a(z) \\ \bar{a}(z) & \bar{b}(z)
\end{array}
\right)
\left[\begin{array}{c} %
\bm{\Psi}_z(t) \\
\bar{\bm{\Psi}}_z(t) \end{array} \right]
\label{linear_1}
\end{eqnarray}
with the $a$, $b$'s being the transmission and reflection
coefficients respectively. By taking into account the
symmetry of the problem under complex conjugation 
it is possible to show that $a(z^*)
=\bar{a}^*(z)$ and $b(z^*) = -\bar{b}^*(z)$, where the star
($^*$) denotes the complex conjugate.

When the above solutions correspond to a localized eigenfunction
with eigenvalue $z$, the transmission coefficient $a(z)$ has to
vanish at that $z$,
making the two sets of solutions directly proportional:
\begin{eqnarray}
\begin{array}{c}
\bm{\Phi}_{z}(t) = b_z\bm{\Psi}_{z}(t) \\ \\
\bar{\bm{\Phi}}_{z}(t) = -b_z^*\bar{\bm{\Psi}}_{z}(t)
\end{array}
\label{linear_2}
\end{eqnarray}
where $\Phi_z$ and $\bar{\Phi}_z$ are the admissible
exponentially decaying eigenfunctions for $Im(z)>0$ and $Im(z)<0$,
respectively. Note that inside the region where $u(t)$ is finite,
they should decay with a Lyapunov exponent $\kappa(z)$, rather
than with $|Im(z)|$ as in (\ref{zs_asymptotic}). The
proportionality constants $b_z$ in (\ref{linear_2}) are not simply
related to the functions $b(z)$ evaluated at the eigenvalue $z$
\cite{AKNS}. It is clear from above that delocalized states can
only exist when $Im(z)=0$.

The proportionality factors $b_z$ and their corresponding
eigenvalues $z$ are very important quantities in the theory of
the Inverse Scattering Transforms: Together with the continuum
delocalized states characterized by $b(z)$, they can completely
reconstruct the original $u(t)$. Therefore, in the context of
information theory, they carry the same information content. 
 In physical terms, the localized
eigenstates of the Zakharov-Shabat problem correspond (through
the IST) to the solitonic excitations in the fiber, while the
continuous spectrum for $Im(z)=0$ gives the radiation
modes, which spread out and decrease in amplitude as the signal
propagates down the optical channel. We will focus on the
localized states, since in the limit $T\rightarrow\infty$ they
correspond to the dominant part of the solution.

Our computation of the DOS  of the problem is based on the
calculation of the Lyapunov exponent $\kappa(\xi,\eta)$, which
then yields the density of states through the generalized Thouless
formula (derived in Appendix A):
\begin{eqnarray}
\rho(\xi,\eta) = \frac{1}{2\pi}\left(
\frac{\partial^2}{\partial\xi^2}+\frac{\partial^2}{\partial\eta^2}\right)\kappa(\xi,\eta)
\label{thoulessformula}
\end{eqnarray}

The (upper) Lyapunov exponent is defined by:
\begin{eqnarray}
\kappa =
\lim_{t\to\infty}\frac{1}{2t}\ln\left(|\psi_1(t)|^2+|\psi_2(t)|^2\right)
\end{eqnarray}
which can also be written as:
\begin{eqnarray}
\kappa = \lim_{t\to\infty}\frac{1}{2t}\int_{0}^{t}
dt'\frac{\partial}{\partial
t'}\ln\left(|\psi_1(t')|^2+|\psi_2(t')|^2\right). \label{eq5}
\end{eqnarray}
Since the system is self-averaging (the evolution of $\psi_1,
\psi_2$ along $t$ is a Markov process), we can exchange the
average over $t$ in (\ref{eq5}) with an average over the Gaussian
ensemble:
\begin{eqnarray} \kappa =
\frac{1}{2}\lim_{t\to\infty}\left\langle \frac{\partial}{\partial
t}\ln\left(|\psi_1|^2+|\psi_2|^2\right) \right\rangle.
\label{lyapunovdef} \end{eqnarray}

This is our starting point for calculating  $\kappa$. From
(\ref{zs}) we find:
\begin{eqnarray} \partial_t\left(|\psi_1|^2+|\psi_2|^2\right) = 2\eta
\left(|\psi_1|^2-|\psi_2|^2\right). \label{lyapunov1}
\end{eqnarray}
Defining the complex variable
$f(t)=\frac{\psi_1(t)}{\psi_2(t)}=e^{w(t)+i\phi(t)}$, with $w\in
(-\infty,\infty)$ and $\phi\in [0,2\pi)$, we can rewrite
(\ref{lyapunov1}) as:
\begin{eqnarray} \partial_t \ln\left(|\psi_1|^2+|\psi_2|^2\right) = 2\eta\tanh{w}.
\label{lyapunov2}
\end{eqnarray} %
We are interested in the long-time behavior of (\ref{lyapunov2}).
For a given $u(t)$, $w(t)$ undergoes constant change at any $t$,
but the probability distribution of its values in the Gaussian
ensemble will tend to a stationary distribution for large $t$. To
see this, we must derive the Fokker-Planck equation for the joint
probability distribution $P(w,\phi;t)$. This is straightforward
for $\delta$-correlated Gaussian potentials, since in this case
$w(t)$ and $\phi(t)$ become Markov
processes \cite{Halperin_1965_DOS}.

\subsection{Circular complex Gaussian potential $u(t)$}
\label{sec:Circularly symmetric Gaussian potential}

\begin{figure}[htbp]
\hspace*{-45pt} \vspace*{5pt}
\scalebox{.3}{\includegraphics*{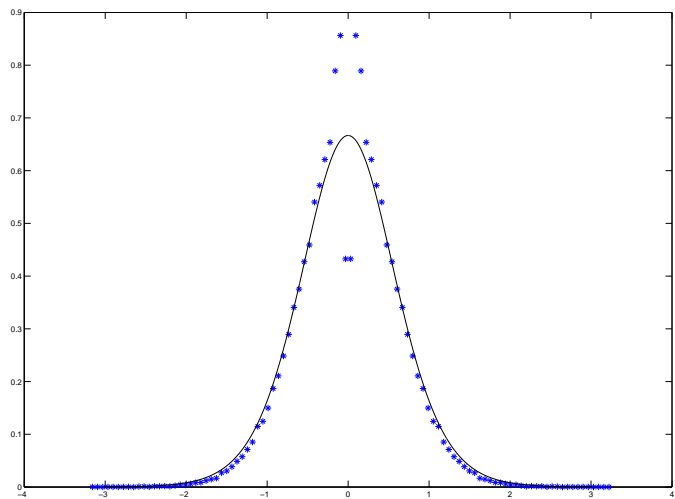}}
\caption{\label{dosfig}Theoretical curve (solid line) and results
of numerical simulations for the profile of the DOS vs. $\eta$. We
have used the modified Ablowitz-Ladik diagonalization scheme
\cite{Weideman_Herbst_Modified_Ablowitz_Ladik} to ensure that the
determinant of the discretized transfer matrix has unit value. The
value of $D$ is $1$, the size of the system is $T=135$ and the
step size is $0.075$. The disturbance near $\eta=0$ is a
finite-size effect. The localization length grows as $l\sim
3D/2\eta^2$ near $\eta=0$ and so numerical results differ from our $T\to \infty$ formula
for $|\eta| \lesssim \sqrt{\frac{D}{T}}$.}
\end{figure}

We start by calculating the density of states (DOS)
$\rho(\xi,\eta)$ and localization length $l(\xi,\eta)$ when $u(t)$
is circularly symmetric, \emph{i.e.}
$u(t)= \frac{1}{\sqrt{2}}\left(u_1(t)+i u_2(t)\right)$
with $u_1,u_2$ real, $\left\langle u_i(t) \right\rangle = 0$, and
$\left\langle
u_i(t)u_j(t')\right\rangle=D\delta_{ij}\delta(t-t')\;$,$\;\;i,j=1,2$.
In this case, the evolution of $w$ and $\phi$ is described by the
set of stochastic equations:
\begin{eqnarray}
\begin{array}{lll}
\partial_t w & = & 2\eta + 2 \cosh{w}\left(\sin{\phi}u_1+\cos{\phi}u_2\right) \\
\partial_t \phi & = & -2\xi - 2\sinh{w}\left(\cos{\phi}u_1-\sin{\phi}u_2
\right).
\end{array}
\label{stochphi}
\end{eqnarray}
The Fokker-Planck equation derived from these (in the Stratonovich
picture) is:
\begin{eqnarray}
\partial_t P & = & \partial_v \left[ (1-v^2)\left( -2\eta + %
D \partial_v \right)\right] P \\ \nonumber %
&+& \partial_\phi\left( 2\xi + D\frac{v^2}{1-v^2}
\partial_\phi\right)P \label{fpcomplexu}
\end{eqnarray}
where $v=\tanh w$. A simplification can be obtained by integrating
over $\phi$. Because the right hand side of (\ref{lyapunov2})
depends only on $v$, we only need $\tilde{P}(v) = \int_{0}^{2\pi}
d\phi P(v,\phi)$ to calculate the average. Integrating over $\phi$
and using the periodicity of $P$ in this variable we find the
Fokker-Planck equation for $\tilde{P}$:
\begin{eqnarray}
\partial_t \tilde{P} & = & \partial_v \left[ (1-v^2)\left( -2\eta + %
D \partial_v \right)\right] \tilde{P}.
\label{fpreduced}
\end{eqnarray}
Setting the left-hand side to zero we find the stationary solution
to which the system relaxes for large $t$:
\begin{eqnarray}
\tilde{P}(v) =
\frac{\eta e^{\frac{2\eta v}{D}}}{D\sinh{\frac{2\eta}{D}}}
\label{stationaryP}
\end{eqnarray}
This is also a stationary solution of the full Fokker-Planck
equation (\ref{fpcomplexu}), implying that asymptotically
$\phi$ becomes uniformly distributed. We can now calculate the
Lyapunov exponent from equations (\ref{lyapunovdef}),
(\ref{lyapunov2}):
\begin{eqnarray}
\kappa  =  \eta \int_{-1}^{1} dv v\tilde{P}(v)
  =  \frac{D}{2} \left( \frac{2\eta}{D}\coth\left( \frac{2\eta}{D} \right) -1 \right) \label{lyapunovcalc}
\end{eqnarray} %
Note that for large $|\eta|$, $\kappa \approx |\eta|$
independently of $D$: this is expected since in this limit the
potential decouples the left ($\psi_1$) from the right
moving ($\psi_2$) wavefunctions. A simple application of
the Thouless formula (\ref{thoulessformula}), gives the exact
density of eigenstates for the system:
\begin{eqnarray}
\rho(\xi,\eta) = \frac{2}{\pi D} \frac{\frac{2\eta}{D}\coth\left(
\frac{2\eta}{D} \right) -1}{\sinh^2\left(\frac{2\eta}{D}\right)}.
\label{dos}
\end{eqnarray}
The independence of $\rho$ from $\xi$ is not surprising: the
density of states of the Hermitian (diagonal) part of (\ref{zs})
is independent of $\xi$. Therefore, in the so-called mean-field
approximation
\cite{Marchetti_Simons_1999_tail_states,Feinberg_Zee_Hermitization}
the extension in the imaginary axis will be $\xi$-independent.
It should be noted however that that mean-field approach
would have given a {\em constant} DOS within a zone around
$\eta=0$, rather than (\ref{dos}). A comparison of this
expression with the result of numerical simulations can be seen in
Fig.~\ref{dosfig}. Again note that for large $\eta$ the density of
states vanishes: in this limit there is an exponentially small
probability for finding a potential deep enough to create a bound
state.

The localization length $l(\xi,\eta)$  is the inverse of the
Lyapunov exponent, $l=\kappa^{-1}$. To see this, we note that the
Wronskian of two independent solutions of (\ref{zs}) is constant,
therefore if for a given $z$ it has an solution increasing
exponentially as $\exp(\kappa t)$, its other solution has to be
exponentially decreasing as $\exp(-\kappa t)$. Thus a square
integrable solution necessarily decays with length-scale
$\kappa^{-1}$ inside the support of $u(t)$. From
(\ref{lyapunovcalc}) one can see that states become increasingly
delocalized as the eigenvalues approach the real axis on the
complex $z$ plane: $l$ diverges as $l\sim \frac{3D}{2\eta^2}$ near
the real axis. The localization length also determines the
stability of the corresponding eigenvalue to the presence of a
finite time window of the pulse $T$. Specifically, the
typical lifetime of a state with eigenvalue $z$ will scale as
$\sim e^{\kappa(\eta)T/2}$ \cite{Gredeskul1990_Dark_Solitons}. 
Indeed we see this in Fig.~\ref{dosfig}, where for states with
localization length comparable to the system length $T$,
\emph{i.e.} close to $\eta=0$, the calculated DOS is no longer
valid. To capture the behavior of the DOS in this region, a
zero-dimensional analysis similar to
\cite{Marchetti_Simons_1999_tail_states,
Efetov_1997_directed_quantum_chaos, Efetov_1997_direction,
Fyodorov_Khor_Sommers_1997}
is needed. %

\subsection{Real initial pulse $u(t)$}
\label{sec:Real potential} %
We can also analyze the opposite case when $u(t)$ is real,
Gaussian with $\left\langle u(t) \right\rangle = 0$, and
$\left\langle u(t)u(t')\right\rangle=D\delta(t-t')\;$. In this
case, the evolution of $w$ and $\phi$ is described by
(\ref{stochphi}) by setting $u_2(t)=Im(u(t)) = 0$. The
corresponding steady state solution of the Fokker-Planck equation
can be derived from:
\begin{eqnarray} \label{fprealu}
0 & = & \partial_v\left[(1-v^2) %
 (\sin^2\phi D\partial_v  -2\eta) \right]P +2\xi\partial_\phi P \\ \nonumber %
&-& \sin2\phi
\partial_\phi\left[\frac{1+v^2}{1-v^2}+2v\partial_v\right]P %
+ \frac{2v^2\cos^2\phi}{1-v^2}  \partial_\phi^2P. %
\end{eqnarray}
For  large $|\xi|$, $P$ is independent of $\phi$ to leading order
in $\xi$. Therefore the large--$\xi$ expansion is essentially
identical to a Fourier expansion. Integrating (\ref{fprealu}) over
$\phi$ gives (\ref{fpreduced}). Thus $\tilde{P}$ is to leading
order identical to that of the circularly symmetric complex $u$.
After some algebra one can derive the next-leading order result.
To order $\mathcal{O}(1/\xi^2)$ the correction to the Lyapunov
exponent is
\begin{eqnarray}
\delta\kappa  = \frac{D^2}{4\xi^2} \left(1-
\frac{y}{D}\coth\left(\frac{2\eta}{D}\right)
+\frac{2\eta^2}{D^2\sinh^2\left(\frac{2\eta}{D}\right)} \right) %
\label{lyapunovcalc_realu}
\end{eqnarray} %
resulting in the following correction to the DOS expression of
(\ref{dos})
\begin{eqnarray}
\delta\rho = \frac{D^2}{\pi\xi^2}
\frac{3\eta\coth\left(\frac{2\eta}{D}\right) -
\frac{6\eta^2}{D\sinh^2\left(\frac{2\eta}{D}\right)} -
\frac{4\eta^2}{D}}{\sinh^2\left(\frac{2\eta}{D}\right)} %
\label{dos_realu}
\end{eqnarray}
In the opposite limit of small $\xi$, we expect the distribution
in $\phi$ to be peaked. Indeed for $\xi=0$, (\ref{fprealu}) has a
solution that is proportional to $\delta(\cos\phi)$. This results
in
\begin{equation}\label{P_w_xi0}
  \tilde{P}(v) = \frac{e^{\eta v /D}}{\pi I_0(\eta/D) \sqrt{1-v^2}}
\end{equation}
with corresponding Lyapunov exponent
\begin{eqnarray}
\kappa(\eta)  =  \frac{\eta I_1(\eta/D)}{I_0(\eta/D)}
\label{lyapunovcalc_realu_xi0}
\end{eqnarray} %
where $I_{0,1}$ are modified Bessel functions of the first kind.
We see that compared to (\ref{stationaryP}), (\ref{P_w_xi0})
is more singular when $|v|\approx 1$, i.e. for large $w$. 

\section{Distribution of $b_z$}

The complex numbers $b_z$, that determine the
asymptotics of the bound states of $\bm U$, can be expressed in
terms of the limiting behavior of the eigenfunctions.
Specifically, for $\eta > 0$ we have from
(\ref{zs_asymptotic},\ref{linear_2})
\begin{eqnarray}
\bm{\Psi}_z(t) \rightarrow \left ( \begin{array}{c} 0 \\ 1 \end{array} \right )
  e^{izt}, \bm{\Psi}_z(-t) \rightarrow \left ( \begin{array}{c} b_z^{-1} \\ 0 \end{array} \right )
  e^{izt} \; \mbox{as} \; t \rightarrow \infty
\end{eqnarray}
Defining $\tilde{\bm{\Psi}}_z(t)\equiv \bm{\Psi}_z(-t)$ we can write:
\begin{eqnarray}
b=\lim_{t\to\infty}b(t),\;\;\; b(t)\equiv\frac{\psi_2(t)}{\tilde{\psi}_1(t)},
\end{eqnarray}
where for convenience we have dropped the subscript $z$.
The time evolution of $\ln{b}$ is found from (\ref{zs}),
\begin{eqnarray}
\frac{\partial\ln{b}}{\partial t} = i \left ( uf + \tilde{u}^* \tilde{f} \right )
\label{lnb_evolution}
\end{eqnarray}
with
$f(t) = \frac{\psi_1}{\psi_2}$, $\tilde{f}(t) =
\frac{\tilde{\psi}_2}{\tilde{\psi}_1}$, $\tilde{u}(t) \equiv
u(-t)$, and
\begin{eqnarray}
\begin{array}{l}
\frac{\partial f}{\partial t} = -2izf + i u^* - iuf^2  \\ \\
\frac{\partial \tilde{f}}{\partial t} = -2iz\tilde{f} - i \tilde{u} + i\tilde{u}^*\tilde{f}^2.
\end{array}
\label{fg_evolution}
\end{eqnarray}

\subsection{Circular complex Gaussian $u(t)$}
For a circularly symmetric complex Gaussian $u$,
the solution of the Fokker-Planck equation derived from (\ref{lnb_evolution},\ref{fg_evolution})
relaxes for large times towards a stationary solution where $\ln{b}$ is uniformly
distributed \footnote{Regarding the initial positions (in time) of the solitons,
this means that the solitons will be uniformly distributed, independently of velocity or amplitude
in the limit $T \rightarrow \infty$. This is expected, given
the translation symmetry of the initial condition.},
while $f$ and $\tilde{f}$, expressed in polar form, \emph{i.e.} $f=e^{w+i\phi}$, $\tilde{f}=e^{\tilde{w}+i\tilde{\phi}}$,
are distributed independently according to the steady state solution (\ref{stationaryP}).
Because of the infinite range of the real part of $\ln{b}$ however,
this stationary solution is ill-defined.
A better approach is to discretize the size $T$ of the pulse into steps of
size $\tau$, equal to the inverse bandwidth of the input signal. Equation (\ref{lnb_evolution})
then reads
\begin{eqnarray}
\ln{b} = i\tau\sum_{i=1}^{T/2\tau}\left( u_i f_i + \tilde{u}_i^* \tilde{f}_i\right).
\label{discrete_lnb}
\end{eqnarray}
The variables $u_i,\tilde{u}_i$ are \emph{i.i.d.} Gaussian random variables,
distributed according to
\begin{eqnarray}
P_u(u) = \frac{1}{\sqrt{2\pi}u_0}e^{-\frac{u^2}{2u_0^2}},
\end{eqnarray}
where $u_0^2 = D/2\tau$ and $u$ stands for the real or imaginary part of either variable.
For large enough $T$, the sum in (\ref{discrete_lnb}) will be
dominated by the domain where the distributions of $f_i$ and $\tilde{f}_i$ have reached
their steady state. In this domain, we find that the
real and imaginary parts of the products $x_i\equiv u_i f_i$ and
$\tilde{x}_i\equiv \tilde{u}_i^* \tilde{f}_i$ have zero mean and the tails of their distributions
fall off as the inverse third power of the argument. More precisely,
\begin{eqnarray}
P(\chi)\sim 4\sqrt{\pi}u_0^2\frac{\eta e^{\frac{2\eta}{D}}}
{D\sinh\left ( \frac{2\eta}{D}\right )}\frac{1}{|\chi|^3}\;,\;\;|\chi|\to \infty
\label{z_distr}
\end{eqnarray}
where $\chi$ stands for the real and imaginary parts of $x_i,\tilde{x}_i$. The general
theory for sums of random variables \cite{feller,gnedenko_kolmogorov,bouchaud} then tells us
that for large $T/\tau$ the distribution of $\ln |b|$ will be Gaussian, with
zero mean and variance
\begin{eqnarray}
\sigma^2 = 4\sqrt{\pi}\frac{\eta e^{\frac{2\eta}{D}}}
{\sinh\left ( \frac{2\eta}{D}\right )}T\ln\frac{T}{2\tau}.
\label{lnb_variance}
\end{eqnarray}
The imaginary part of $\ln b$ is an angle and so, although it
follows the same distribution as the real part, will due to
periodicity become uniformly distributed in $[0,2\pi)$. As
seen in (\ref{linear_2}) for $\eta<0$ the corresponding $b_z$ is
replaced by $-b_z^*$ \cite{AKNS}. Thus their distribution will be
the same as that of the $b$'s, with $\eta$ replaced by $-\eta$ in
(\ref{lnb_variance}).

\subsection{Real Gaussian $u(t)$}
In this case, the Fokker-Planck equation for $\ln b$, derived from
(\ref{lnb_evolution},\ref{fg_evolution}) after setting $u_2$ and
$\tilde{u}_2$ to zero, again predicts that its distribution
becomes uniform as the duration $T$ of the pulse grows to
infinity. Equation (\ref{lnb_evolution}) can be written as:
\begin{eqnarray}
\begin{array}{lll}
\frac{\partial\ln{b}}{\partial t}& = &-\left( ue^w\sin{\phi}+\tilde{u}e^{\tilde{w}}\sin{\tilde{\phi}} \right ) \\
&& +i\left(ue^w\cos{\phi}+\tilde{u}e^{\tilde{w}}\cos{\tilde{\phi}} \right )
\end{array}
\label{lnb_evolution_real_u}
\end{eqnarray}
As seen above, an exact solution to (\ref{fprealu}) is not
available, but we can still obtain the first terms of an expansion
of the stationary probability distribution in powers of $e^{-w}$:
\begin{eqnarray}
P(w,\phi) \approx
\alpha_1e^{-w}\delta(\cos{\phi})+\alpha_2e^{-2w}+\mathcal{O}\left(
e^{-3w} \right) \label{large_w_expansion}
\end{eqnarray}
with an identical expansion for the distribution of $\tilde{w}$
and $\tilde{\phi}$. The constants $\alpha_1, \alpha_2$ depend on
$\xi/D$ and $\eta/D$, but, being related to the normalization,
they cannot be determined without a knowledge of the full
solution. We can thus only partially specify the manner in which
the real and imaginary parts of $\ln{b}$ approach uniformity as
$T$ grows.

As in the complex case, the real part, $\ln{|b|}$, will be
a sum of independent variables $\chi_i=u_ie^{w_i}sin\phi_i$.
However, in this case, due to the more singular behavior of
$P(w,\phi)$ for large $w$, the tails of $\chi_i$ will be longer,
falling off as $1/|\chi_i|^2$ for large $T$. As a result,
the distribution of $\ln|b|$ will asymptotically follow a
Cauchy distribution scaling like $T/\tau$
\cite{gnedenko_kolmogorov,
bouchaud}. Its statistical median
will be zero by symmetry, coming from the even parity of the
Gaussian distribution of $u,\tilde{u}$. The phase of $b$ does not
get contributions from the first term in (\ref{large_w_expansion})
because of the delta function in this term. For large $T$, the
second term in the expansion dominates, making it uniform over
$[0,2\pi)$, in the same manner we saw in the case of complex $u$.
Note that for the special case of $\xi=0$, the exact solution (cf.
(\ref{P_w_xi0})) is proportional to $\delta(\cos\phi)$. The scale
parameter of the Cauchy distribution will be
\begin{equation}
\gamma \sim \frac{e^{\eta /D}}{I_0(\eta/D)}\frac{T}{\tau}.
\end{equation}
Only the transients of the distribution add to the phase of $b$,
and numerical simulation shows that they are enough to again make
it uniform.

\section{Discussion}

In the context of the NLSE, the scattering data of the ZS operator
uniquely determine the solitonic excitations we get in the optical
fiber if we feed one end with a delta-correlated Gaussian signal.
Even though the informational contents of the Gaussian signal and
its solitonic spectrum are the same, it is easier to consider the
effect of amplifier noise in the domain of the scattering data.
For example, a small amount of amplifier noise will randomly shift
each eigenvalue $z$ by a small amount, while making large changes
in the output signal \cite{Konotop_book}. The effect of this noise
is important to analyze, in order to calculate the ultimate
information capacity limits through optical fibers. In principle,
to find the capacity one needs to optimize over input signal
distributions, which is a formidable task. Instead, in this paper
we start with a given input distribution and calculate the
corresponding density of states and the corresponding
distribution of scattering data $b_z$. We leave the analysis of
the effects of noise on the spectrum for a future publication.

\begin{acknowledgments}

We would like to thank A. M. Sengupta for many useful discussions in the beginning of this work.

\end{acknowledgments}

\appendix
\section{Thouless Formula}

The proof of the Thouless formula for the ZS eigenproblem proceeds
similarly to the proofs in \cite{Thouless_1972,Hayn_John_1987}. We
consider the system of equations (\ref{zs}) on the interval
$[0,T]$. Let $\bm{\Psi}^l(t)$,$\bm{\Psi}^r(t)$ be two independent
solutions of (\ref{zs}) that satisfy the conditions:
\begin{eqnarray}
\bm{\Psi}^l(0)=\left( \begin{array}{c}
\psi^l_1(0) \\ \psi^l_2(0)
\end{array} \right)\;,\;\;\bm{\Psi}^r(T)=\left( \begin{array}{c}
\psi^r_1(T) \\ \psi^r_2(T) \end{array} \right). \label{initcond}
\end{eqnarray}
We will need to combine this pair of initial and final conditions
into a set of boundary conditions for the eigenstates, and for
this we let each of them be a one-parameter family of
initial(final) conditions to avoid over-determining the problem.
This means that $\psi^l_{1,2}(0)$ are not chosen independently,
but satisfy a single linear relation. The same goes for
$\psi^r_{1,2}(T)$. The Wronskian of the two solutions,
$W=\psi^r_1\psi^l_2 - \psi^r_2\psi^l_1$ is constant. Taking the
derivative of (\ref{zs}) with respect to $z$, we obtain an
equation for $\partial_z\bm{\Psi}^l(t)$ whose solution can be
written in terms of a matrix Green function $\bm{G}^l$:
\begin{eqnarray}
\frac{\partial}{\partial
z}\bm{\Psi}^l(t) = - \int_{0}^{t} dt'
\bm{G}^l(t,t')\bm{\Psi}^l(t') \label{intpsil}
\end{eqnarray}
with
\begin{eqnarray}
\bm{G}^l(t,t') = \left\{
\begin{array}{ll}
\frac{i}{W}\left(\bm{\Psi}^r(t)\left(\bm{\sigma}_1\bm{\Psi}^l(t')\right)^T
\right.&   \\\left. \;\;\;\;\;\;\;\;\;-
\bm{\Psi}^l(t)\left(\bm{\sigma}_1\bm{\Psi}^r(t')\right)^T\right) &
,\;t>t' \\ 0 & ,\;t<t' \end{array} \right.
\end{eqnarray}
Here
$\bm{\sigma}_1 = \left(\begin{array}{lr} 0 & 1 \\ 1 & 0
\end{array} \right)$. The matrix Green function $\bm{G}^l$
satisfies the initial conditions
\begin{eqnarray} \bm{G}^l(0,t') =
0\;,\;\; \left.\frac{\partial}{\partial
t}\bm{G}^l(t,t')\right|_{t=0} = 0 .
\end{eqnarray}
We also define
another Green function,
\begin{eqnarray} \bm{G}(t,t') = \left\{
\begin{array}{l}
\frac{i}{W}\bm{\Psi}^r(t)\left(\bm{\sigma}_1\bm{\Psi}^l(t')\right)^T,\;
t>t'  \\
\frac{i}{W}\bm{\Psi}^l(t)\left(\bm{\sigma}_1\bm{\Psi}^r(t')\right)^T
,\; t<t' \end{array}  \right.
\end{eqnarray}
which satisfies the conditions (\ref{initcond}) (taken together as
boundary conditions) and will determine the density of states. The
Lyapunov exponent can be expressed as\footnote{The average over
the Gaussian enesmble is included here so that our final formula
refers to the average DOS. It is not needed for the definition of
$\kappa$.}:
\begin{eqnarray}
\kappa = \lim_{T\to\infty}\frac{1}{2T}\left\langle
\ln\left({\bm{\psi}^l}^{\dagger}(T)\bm{\psi}^l(T)\right)\right\rangle.
\label{lyapunovdefinition}
\end{eqnarray}
Before going any further, we must note that the value of $\kappa$
is, with probability one, independent of the initial conditions
satisfied by $\bm{\Psi}^l$ (the argument is very similar to that
for the FGM \cite{Bartosch_thesis}). To see this, we rewrite the
system of equations (\ref{zs}) as:
\begin{eqnarray}
i\frac{\partial}{\partial t}\left( \begin{array}{r} \psi_1 \\ \psi_2  \end{array} \right) =
\bm{V} \left( \begin{array}{r} \psi_1 \\ \psi_2  \end{array}
\right)\;,\;\; \bm{V} = \left( \begin{array}{lr} z & -u^* \\ -u &
-z \end{array} \right). \end{eqnarray} We can formally write the
solution for $\bm{\Psi}^l$:
\begin{eqnarray}
\bm{\Psi}^l(t)=\bm{S}(t,0)\bm{\Psi}^l(0)
\\ \bm{S}(t,0)=\mathcal{T}\exp\left[ -i\int_{0}^{t}dt'\bm{V}(t')
\right]
\end{eqnarray}
where `$\mathcal{T}\exp$' denotes the path-ordered exponential
\footnote{Note that $\bm{V}$ is Hermitian (and $\bm{S}(t,0)$ is
unitary) when $z$ is real, and so the Lyapunov exponent is always
zero in that case, irrespective of $u$. Therefore, any
eigenstates on the real axis are {\em extended}, corresponding to scattering
states of the ZS eigenproblem.}. Because the trace of
$\bm{V}$ vanishes, we have $\det{\bm{S}}=1$
\cite{Grensing_path_ordered_exponential}. Therefore, if we denote
the two eigenvalues of $\bm{S}(t,0)$ by $s_{\pm}(t)$, with
$|s_{+}(t)|>|s_{-}(t)|=|s_{+}(t)|^{-1}$, we have
\begin{eqnarray}
\kappa  & = &  \lim_{T\to\infty} \frac{1}{2T}\ln \left(
|s_+(T)\tilde{\psi}_+(T)|^2
+ |s_-(T)\tilde{\psi}_-(T)|^2 \right) \nonumber \\
& = & \lim_{T\to\infty} \frac{1}{T}\ln |s_+(T)|,
\end{eqnarray}
independent of the initial condition, given that the coefficient
of the exponentially increasing solution does not vanish, which is
the case with probability one in the limit of large $T$.

Now, since ${\bm{\Psi}^l}^{\dagger}(T)$ depends only on $z^*$ and
not on $z$, taking the derivative of (\ref{lyapunovdefinition})
with respect to the latter we get
\begin{eqnarray}
\frac{\partial \kappa}{\partial z} = \lim_{T\to\infty}
\frac{1}{2T}\left\langle\frac{{\bm{\Psi}^l}^{\dagger}(T)\frac{\partial\bm{\Psi}^l(T)}{\partial
z}}{{\bm{\Psi}^l}^{\dagger}(T)\bm{\Psi}^l(T)}\right\rangle.
\end{eqnarray}
The quantity inside the average can be computed
from (\ref{intpsil}):
\begin{eqnarray}
\frac{{\bm{\Psi}^l}^{\dagger}(T)\partial_z\bm{\Psi}^l(T)}{{\bm{\Psi}^l}^{\dagger}(T)\bm{\Psi}^l(T)}
& = & -i\frac{{\bm{\Psi}^l}^{\dagger}(T)\bm{\Psi}^r(T)}{W}
\frac{\int_{0}^{T} dt'\left(2\psi^l_1(t')\psi^l_2(t')\right)}{{\bm{\Psi}^l}^{\dagger}(T)\bm{\Psi}^l(T)} \nonumber \\
&& \nonumber \\
 & & + \int_{0}^{T} dt'\mbox{Tr}\bm{G}(t',t')  \label{quaninbrack}
\end{eqnarray}
The first term on the right hand side of (\ref{quaninbrack})
is almost surely  $\mathcal{O}(1)$ in the limit of large $T$ and so
does not contribute to the average. We are thus left with
\begin{eqnarray}
\frac{\partial \kappa}{\partial z} =
\lim_{T\to\infty}\frac{1}{2T}\left\langle \int_{0}^{T}
dt'\mbox{Tr}\bm{G}(t',t') \right\rangle \label{lambdader}
\end{eqnarray}

Taking the derivative of (\ref{lambdader}) with respect to $z^*$
and using the relation\cite{Feinberg_Zee_Hermitization}
$\rho(z,z^*) = \frac{1}{\pi T}\frac{\partial}{\partial
z^*}\left\langle \mbox{Tr}\bm{G} \right\rangle$ we arrive at the
Thouless formula for the density of states:
\begin{eqnarray}
\rho(z,z^*)=\frac{2}{\pi}\frac{\partial^2\kappa}{\partial z^*
\partial z}=\frac{1}{2\pi}\left(
\frac{\partial^2}{\partial\xi^2}+\frac{\partial^2}{\partial\eta^2}\right)\kappa.
\end{eqnarray}

%\bibliographystyle{apsrev}
%\bibliography{wireless}
%\bibliography{pavlosref}
%\bibliographystyle{IEEEtran}
%\bibliography{IEEEfull,../bibliography/wireless}

\end{document}